\documentclass[]{ws-ijmpb}
\usepackage{amsmath,amstext,amssymb,graphics,setspace,bm,dcolumn,color,psfrag}

\def\bra#1{\mathinner{\langle{#1}|}}
\def\ket#1{\mathinner{|{#1}\rangle}}
\def\braket#1{\mathinner{\langle{#1}\rangle}}

{\catcode`\|=\active 
 \gdef\Braket#1{\begingroup
\mathcode`\|32768\let|\BraVert\left<{#1}\right>\endgroup}}
\def\BraVert{\egroup\,\mid\,\bgroup}

\definecolor{Blue}{rgb}{0,0,1}
\definecolor{Red}{rgb}{1,0,0}
\definecolor{Green}{rgb}{0,1,0}
\definecolor{Purp}{rgb}{.2,0,.2}
\definecolor{white}{rgb}{1,1,1}

\begin{document}

\markboth{Kavan Modi and Mile Gu}
{Coherent and incoherent contents of correlations}

%%%%%%%%%%%%%%%%%%%%% Publisher's Area please ignore %%%%%%%%%%%%%%%
%
\catchline{}{}{}{}{}
%
%%%%%%%%%%%%%%%%%%%%%%%%%%%%%%%%%%%%%%%%%%%%%%%%%%%%%%%%%%%%%%%%%%%%

\title{Coherent and incoherent contents of correlations}

\author{Kavan Modi$^{1,2,}$\footnote{Email: kavan@quantumlah.org} and Mile Gu$^{2,}$\footnote{Email: ceptryn@gmail.com}}
\address{$^1$Clarendon Laboratory, Department of Physics, University of Oxford, Oxford, UK and\\$^2$Centre for Quantum Technologies, 
National University of Singapore, Singapore}

\maketitle

\begin{abstract}
We examine bipartite and multipartite correlations within the construct of unitary orbits. We show that the set of product states is a very small subset of set of all possible states, while all unitary orbits contain \emph{classically correlated} states. Using this we give meaning to degeneration of quantum correlations due to a unitary interactions, which we call coherent correlations. The remaining classical correlations are called incoherent correlations and quantified in terms of the distance of the joint probability distributions to its marginals. Finally, we look at how entanglement looks in this picture for the two-qubit case.
\end{abstract}


%%%%%%%%%%%%%%%%%%%%%%%%%%%%%%%%%%%%%%%%
%%%%%%%%%%%%%%%%%%%%%%%%%%%%%%%%%%%%%%%%
\section{Introduction}
%%%%%%%%%%%%%%%%%%%%%%%%%%%%%%%%%%%%%%%%
%%%%%%%%%%%%%%%%%%%%%%%%%%%%%%%%%%%%%%%%

The studies of quantum correlations beyond entanglement, specifically discord and similar measures, have exploded in many different contexts over the last five years\cite{arXiv:1112.6238}. A goal of these studies is often to separate quantum and classical parts of correlations\cite{henderson01a}. Indeed this was the motivation in the very discovery of quantum discord\cite{PhysRevLett.88.017901}, deficit\cite{arXiv:quant-ph/0112074,arXiv:0911.5417}, measurement induced disturbance\cite{PhysRevA.66.022104,PhysRevA.77.022301}. What all of these measures have in common are the classical states.\cite{arXiv:1002.4913,arXiv:1004.0190,arXiv:1005.4348}, and perhaps it is this boundary of quantum and classically correlated states that is most important\cite{CesarEtal07,arXiv:0707.0848}. This boundary is different from the boundary of entangled states and separable states\cite{PhysRevA.40.2477,arXiv:0707.2195}. The boundary arises naturally as the states that are not disturbed by the measurements process are called classical. However, making measurements is not the only way to attain classical states. 

In this article we define quantum in terms of global unitary operations. This is to be interpreted as the amount of coherent interaction needed to extract all information contained in a multipartite state. This is different from a recently introduced measure of quantum correlations in terms of local unitary operations\cite{arXiv:1202.1598}. We begin by defining the notion of unitary orbits. Then we show that almost all unitary orbits do not contain uncorrelated states and therefore classical correlations are unavoidable, when dealing with mixed states. Next, we present the promised measure of quantum correlations, and conclude by discussing the result in context of a recent theory and experiment\cite{arXiv:1203.0011} on consumption of discord as a resource.

%%%%%%%%%%%%%%%%%%%%%%%%%%%%%%%%%%%%%%%%
%%%%%%%%%%%%%%%%%%%%%%%%%%%%%%%%%%%%%%%%
\section{Unitary orbits}
%%%%%%%%%%%%%%%%%%%%%%%%%%%%%%%%%%%%%%%%
%%%%%%%%%%%%%%%%%%%%%%%%%%%%%%%%%%%%%%%%

The geometry of quantum states is an old and a rich topic\cite{geobook}. In this paper, we explore the space of unitary orbits, which we define following Boya and Dixit\cite{boyadixit}.

{\bf Definition.} A unitary orbit is a set of state that are connected by unitary operations:
\begin{gather}
\varrho\equiv \{U\rho\;U^\dag\} \;\; \forall\;\; U,
\end{gather}
where $\rho$ is a density matrix and $U$ is any unitary operator. $\square$ %in the same space, and $\varrho$ represents the set of all states that are connected by all possible unitary operations; hence we call it \emph{a unitary orbit}.

A unitary transformation is a transformation of basis, which leaves the spectrum unchanged. Any two density matrices that share a spectrum, belong to a single unitary orbit; conversely if 
\begin{gather}
\rho_1\in \varrho \quad {\rm and} \quad \rho_2\in \varrho, \quad {\rm then} \quad Spec(\rho_1)=Spec(\rho_2).
\end{gather}
 Therefore a unitary orbit is uniquely identified by the spectrum of the states it contains. Consequently, all states on a unitary orbit will also have the same scalar measures such as the von Neumann entropy and purity. On the other hand, the analysis here has more depth than a study of scalar measures alone because \emph{two different} unitary orbits can share a single value for \emph{entropy} or \emph{purity}\cite{boyadixit}. It may be tempting to simply study the space of the eigenvalues. In fact, that is all we are doing, but we are taking advantage of the nice properties of unitary transformations to connect various density states.

\begin{figure}[t!]
\begin{center}
\resizebox{8.25 cm}{4.5 cm}
{\includegraphics{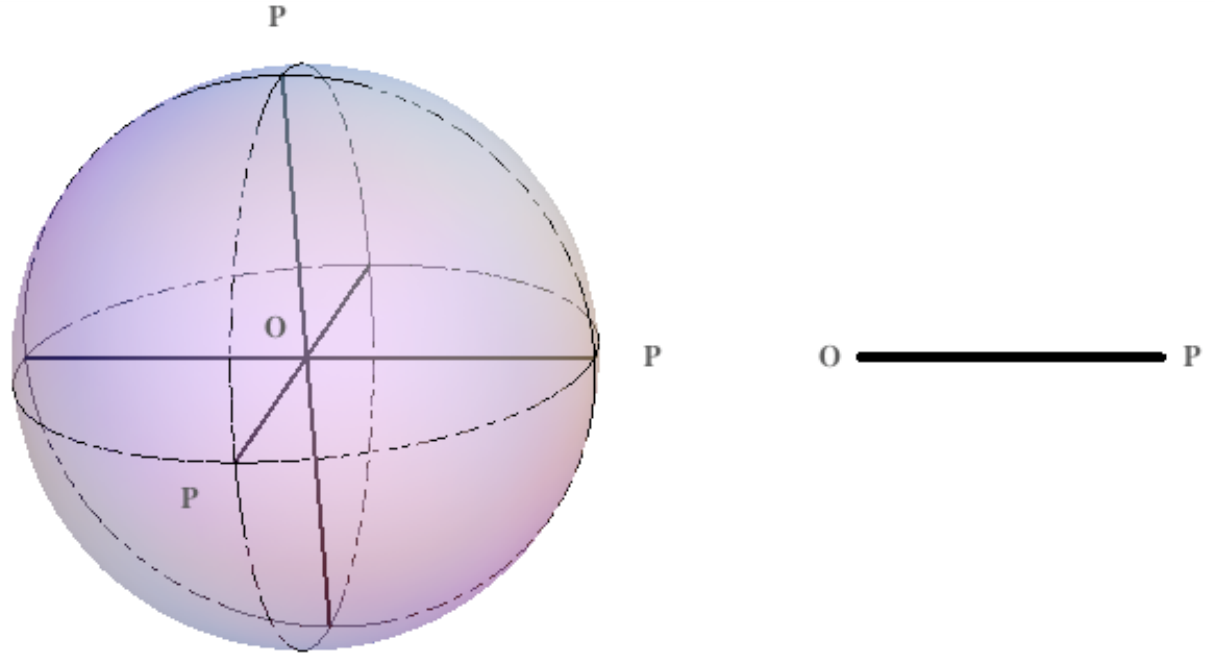}}
\caption{\label{blochsphere}(Color Online.) The states of a qubit are represented by the points inside the Bloch sphere (left). The set of states on each spherical shell are connected by a unitary transformation, therefore the sphere goes into a line (right) when the set of all qubit states are mapped to set of unitary orbits of a qubit. $P$ represents the set of all pure states and $O$ represents the fully mixed state.}
\end{center}
\end{figure}

Let us now briefly discuss the familiar example of the space of one qubit represented by the points inside of the Bloch sphere, shown on the left in Fig.~\ref{blochsphere}. The outer most shell of the Bloch sphere represents the set pure states, the centre point represents the fully mixed state, and everything else is represented by the space inside of the sphere. The states on each of the spherical shells of the Bloch sphere are connected to each other via unitary transformations. To construct the geometry of unitary orbits, we contract each shell of the Bloch sphere into a single point. That is, all of the states that are unitarily connected are now represented by an \emph{unitary orbit}. The geometry for the unitary orbit, for a qubit, simplifies to line, shown on the right in Fig.~\ref{blochsphere}. The two end points of the line are the fully mixed state labeled as $O$ and the set of all pure states labeled as $P$.

One can carry out such a procedure for any dimensional systems. For a qutrit, a three-level system, the geometry of states is given by an eight dimensional manifold. The geometry of unitary orbit, however, is given by a triangle\cite{boyadixit}. In general the manifold that represents the set of state for a $d-$dimensional system grows as $d^2-1$, while the manifold of unitary orbit grows linearly $d-1$. Furthermore the manifold for unitary orbits is a simplex with $d$ vertices. One can think of each unitary orbit as a $d+1$ dimensional manifold contracted to a point.

%%%%%%%%%%%%%%%%%%%%%%%%%%%%%%%%%%%%%%%%
%%%%%%%%%%%%%%%%%%%%%%%%%%%%%%%%%%%%%%%%
\section{Product States}
%%%%%%%%%%%%%%%%%%%%%%%%%%%%%%%%%%%%%%%%
%%%%%%%%%%%%%%%%%%%%%%%%%%%%%%%%%%%%%%%%

In this paper we want to analyse the space of unitary orbits for composite systems with three different cases in mind. First, we want look at the unitary orbits of a composite system that contain product states, followed by classically correlated states, and finally entanglement within unitary orbits. Before we proceed, we should remark that generally in quantum information theory one is interested in local operations\cite{Book.Nielsen.Chuang}. The unitary orbits of composite system that we consider here are due to the actions of global unitary transformations. Let us begin with examining how the product state lie in a unitary orbit.

{\bf Theorem.} The subspace of unitary orbits containing product states of $n$ parties is a $\sum_i d_i-n$ dimensional surface in a $\prod_i d_i-1$ dimensional manifold, where $d_i$ is the dimensions of $i$th subsystem.

\emph{Proof.} Suppose that a unitary orbit, $\varrho_p$, contains a $n-$partite product state 
\begin{gather}
\pi= \bigotimes_i^n \pi^{(i)}
\equiv \pi^{(1)} \otimes \pi^{(2)} \otimes \pi^{(3)} \cdots \otimes \pi^{(n)}.
\end{gather}
The spectrum of $\pi$ is simply the outer product of the spectrums of each of the subsystem and any state $\rho\in\varrho_p$ has the same spectrum. Conversely, any unitary orbit that contains a product states has a spectrum that is factorable in terms of the spectrum of the subparts. Each of the subparts above has $d_i-1$ independent parameters (eigenvalues); the total number of independent parameters for $\varrho_p$ is simply the sum of independent parameters of each of the subpart, $\sum_i^n (d_i-1)=\sum_i^n d_i-n$. While the number of independent parameters for a generic unitary orbit is given by $\prod_i^n d_i-1$. $\square$

Let us illustrate the consequences of the theorem with a simple example of a two-qubit composite system. But first, a word of caution is necessary. If a unitary orbit contains a product state that \emph{does not} mean that all states on the that orbit are of the product form. A simple example is a maximally entangled state: it is a pure state belonging to the unitary orbit of pure states, yet it is not of product form. But, of course, the unitary orbit containing the maximally entangled state will also contain pure product states.

%%%%%%%%%%%%%%%%%%%%%%%%%%%%%%%%%%%%%%%%
\subsection{The two qubit case}
%%%%%%%%%%%%%%%%%%%%%%%%%%%%%%%%%%%%%%%%

%%%%%%%%%%%%%%%%%%%%%%%%%%%%%%%%%%%%%%%%
\begin{figure}[t!]
\begin{center}
\resizebox{8.54cm}{4.5cm}
{\includegraphics{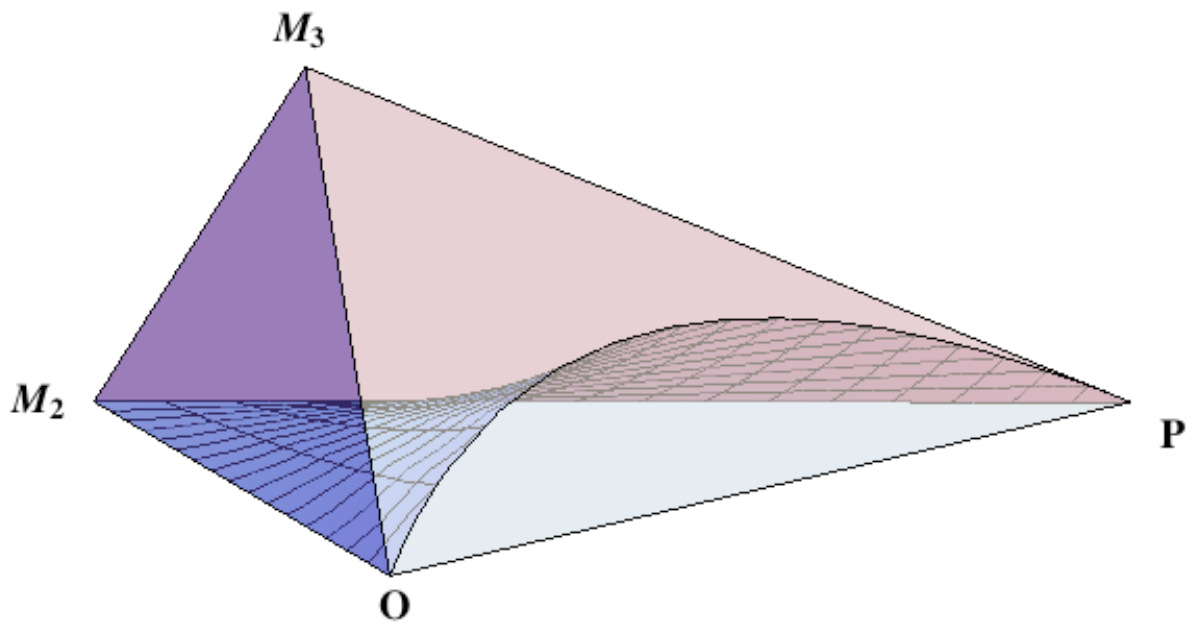}}
\caption{\label{productspace}(Color Online.) The geometry of unitary orbits for two qubit is represented by a tetrahedron. Above, $P$ represents all pure states, $M_2$ represents states with two identical non-zero eigenvalues, $M_3$ represents states with three identical non-zero eigenvalues, and $O$ represent the fully mixed state. The surface through the tetrahedron represents the set of unitary orbits that contain product states.}
\end{center}
\end{figure}
%%%%%%%%%%%%%%%%%%%%%%%%%%%%%%%%%%%%%%%%

The smallest composite system is the two-qubit system with $d=4$. The space of the unitary orbits for the two-qubit case is confined to a tetrahedron the three dimensional space\cite{boyadixit}. We are interested to identify the region of the tetrahedron that is occupied by unitary orbits that contain product states. To solve this problem we need to match the eigenvalues of a generic two-qubit state to the eigenvalues of a qubit state in outer product with itself
\begin{eqnarray}\label{const1}
&&	Spec\left(\rho^A\right)\times Spec\left(\rho^B\right)
=Spec\left(\rho^{AB}\right)\nonumber\\
&&	\left\{\begin{matrix}
		1+a+b+ab\cr
		1+a-b-ab\cr
		1-a+b-ab\cr
		1-a-b+ab\cr
	\end{matrix}\right\}
=	\left\{\begin{matrix}
			1+\sqrt{6}x+\sqrt{2}y+z\cr
			1-\sqrt{6}x+\sqrt{2}y+z\cr
			1-2\sqrt{2}y+z\cr
			1-3z\cr
		\end{matrix}\right\}.
\end{eqnarray}
Above the spectrum of a two-qubit states is given in terms of three parameters as\cite{boyadixit,byrd062322}. The constraint of Eq.~\eqref{const1} leads to equation 
\begin{gather}
z^2 + z + \frac{1}{\sqrt{2}}(z-1) \left(\sqrt{3} x +y \right) +\sqrt{3} xy - y^2=0
\end{gather}
with two solutions for $z$. We only consider the positive solution, plotted in Fig.~\ref{productspace}. Note that we have made one additional contraction by ordering the eigenvalues from large to small (top to bottom in Eq.~\eqref{const1}). In that sense, we have contracted twenty-four tetrahedrons (permutations of four) into one tetrahedron\cite{boyadixit}. The negative solution therefore belongs to a space that has been contracted by permutation symmetry. The solution of positive $z$ shows that the set of unitary orbits containing products states are confined to a two-dimensional plane in a three-dimensional manifold.

%%%%%%%%%%%%%%%%%%%%%%%%%%%%%%%%%%%%%%%%
\subsection{Higher dimensions and other considerations}
%%%%%%%%%%%%%%%%%%%%%%%%%%%%%%%%%%%%%%%%

Now we are in the position to make a few remarks about unitary orbits with product states embedded within the space all unitary orbits. Once again, the dimensions for the unitary orbit manifold is given by the product of dimensions of each subspace, while the dimensions of the manifold that represent the unitary orbits that contain product states is given by the sum of the number of independent parameters of each subspace. The surface of unitary orbits containing product states for $d=6$ (a qubit-qutrit system) is a $3-$dimensional surface in a $5-$dimensional volume. For $d=8$ (three qubit system) case with, the surface of unitary orbits containing product states is also a $3-$dimensional surface, but in a $7-$dimensional volume. The set of correlated states in the three-qubit system is much higher than the correlations in the qubit-qutrit system for product states as the base measure. The space of unitary orbits grows much faster than the subspace that contains product states; this should be cautionary statement for an analysis that depends on only product states; as it may lack generality. Let us divert the discussion for the moment an examine such a situation next.

%%%%%%%%%%%%%%%%%%%%%%%%%%%%%%%%%%%%%%%%
\subsection{Consequences for open systems}
%%%%%%%%%%%%%%%%%%%%%%%%%%%%%%%%%%%%%%%%

We showed that the set of unitary orbits containing product states is a very small subset of set of all unitary orbits. Which in return means that the set product states is a very small subset of set of all states. Any realistic model of a multipartite quantum system \emph{should not} assume that the parts are simplify separable. That assumption leads to gross simplification which could undermine a great deal of physical effects. There has been a great deal of discussion regarding initially product state assumption in open quantum dynamics\cite{PhysRevLett.75.3020,pechukas94a,jordan052110,CesarEtal07}. We have now definitively shown that when initially product state assumption is retained one is restricted to a small subspace of the total state space. An immediate application of the calculations above is seen in the derivation of the non-Markovian master equation due to Rodr\'iguez-Rosario and Sudarshan\cite{cesarnonmarkov}. In their derivation they start with a generic state of the system and the environment, which is assumed to be of product form at some point in its (unitary) history, which is not a good assumption for a generic state.

%%%%%%%%%%%%%%%%%%%%%%%%%%%%%%%%%%%%%%%%
%%%%%%%%%%%%%%%%%%%%%%%%%%%%%%%%%%%%%%%%
\section{Classical correlated states}
%%%%%%%%%%%%%%%%%%%%%%%%%%%%%%%%%%%%%%%%
%%%%%%%%%%%%%%%%%%%%%%%%%%%%%%%%%%%%%%%%

Next let us look at the case of classically correlated states (not separable states), as defined by Henderson and Vedral\cite{henderson01a}, and Ollivier and Zurek\cite{PhysRevLett.88.017901}.

{\bf Definition.} A multipartite state is called classical if it has the form:
\begin{gather}
\chi_{B|A} = \sum_a^{d_A} p_a \ket{a}\bra{a}\otimes \pi_{B|a},
\end{gather}
where $\{\ket{a}\}$ form an orthonormal basis in the space of $A$,
$p_a$ are classical weights satisfying $\sum_a p_a=1$, and $d_A$ is dimensions of subsystems $A$. The state above is called one-way classically correlated state. A state is called fully-classical is it is classical for both $A$ and $B$
\begin{gather}
\chi_{AB} = \sum_{a}^{d_A}\sum_{b}^{d_B} p_{ab} \ket{a}\bra{a}\otimes\ket{b}\bra{b}.
\end{gather}
These states are said to posses no quantum correlations, as they simply mimic classical joint probability distributions\cite{arXiv:0707.0848}.

{\bf Theorem.} All unitary orbits contain one-way and fully-classically correlated states.

\emph{Proof.} Let us start with one-way classically correlated states, which are block diagonal. Consider a set of unitary operators in the space of $B$ that diagonalise the $a$th state, $\pi_{B|a}\rightarrow u^{(a)}_B \pi_{B|a} {u^{(a)}_B}^\dag = \sum_{b} p_{b|a} \ket{b}\bra{b},$ where $p_{ab}$ is the $b$th eigenvalues of $\pi_{B|a}$. Let us now construct the following \emph{control-unitary} operator $U^c=\sum_{a}\ket{a}\bra{a}\otimes u^{(a)}_B$. Applying $U^c$ to $\chi_{B|A}$ yields a fully classical state.
\begin{align}
\chi_{AB}
=&U^c \chi_{B|A} {U^c}^\dag\\
=&\sum_{ab} p_{ab} \ket{a}\bra{a} \otimes \ket{b}\bra{b}.
\end{align}
Above $p_{ab}=p_a p_{b|a}$ are the eigenvalues of state $\chi_{A|B}$. This shows that all one-way classically correlated states are unitarily connected to fully classically correlated states.

Next, consider an arbitrary state 
\begin{gather}
\rho =\sum_{ab} p_{ab} \ket{v_{ab}} \bra{v_{ab}}, 
\end{gather}
with eigenvalues $p_{ab}$ and (generally not separable) basis $\ket{v_{ab}}$, where $a$ and $b$ run up to the dimensions of spaces $A$ and $B$ respectively. By the definition of a unitary transformation there exists an operator $W$ that connects the basis $\{\ket{ab}\}$ to the basis $\{\ket{v_{ab}}\}$, as $W\ket{v_{ab}}=\ket{ab}$. Next, rewrite the eigenvalues $p_{ab}$ as $p_{ab}= p_a p_{b|a}$, where $p_a \equiv \sum_a p_{ab}$ and $p_{b|a} \equiv \frac{p_{ab}}{p_a}$. By definition both $p_a$ and $p_{b|a}$ are positive numbers less than (or equal to) 1 satisfying the conditions 
\begin{gather}
\sum_b p_{b|a} =\sum_{b} \frac{p_{ab}} {p_a} =\frac{\sum_{b}p_{ab}} {\sum_{b}p_{ab}}=1
\quad {\rm and} \quad 
\sum_a p_a=\sum_{ab}p_{ab}=1.
\end{gather} 

Finally, we can connect a generic state $\rho$ to a fully-classically correlated state $\chi_{AB}$ by acting with $W$:
\begin{align} 
W \rho W^\dag \nonumber
=&\left(\sum_{ab} p_a p_{b|a}
\ket{ab}\bra{ab}\right) = \chi_{AB}.
\end{align}  
Since we already showed that fully-classical states are connected to one-way classical states, this completes the proof. $\square$

There is no real difference between the space $AB$ and space $BA$ apart from labelling; the result above can be extended for classical correlated states in the space of $B$ by swapping the definitions for $p_a$ and $p_{b|a}$. Note that if an $p_{a=k}=0$, that means that all $p_{kb}=0$ for all $j$. The only requirement we have so far set is that $p_{ab} = p_a p_{b|a}$. When the left hand side is zero, $p_a$ on the right hand side is also zero, and $p_{b|a}$ can be anything. Another way to look at this is by considering $p_a$ as classical wights. When the classical weight is zero then the corresponding term for the subsystem $B$ can be anything or neglected all together. The proof above can be generalised to a multipartite scenario by considering all but one system together, and then the same procedure can be repeated. 

%%%%%%%%%%%%%%%%%%%%%%%%%%%%%%%%%%%%%%%%
%%%%%%%%%%%%%%%%%%%%%%%%%%%%%%%%%%%%%%%%
\section{Quantifying correlations}
%%%%%%%%%%%%%%%%%%%%%%%%%%%%%%%%%%%%%%%%
%%%%%%%%%%%%%%%%%%%%%%%%%%%%%%%%%%%%%%%%

We have now shown that not all mixed states can be unitarily turned into product states. However, they may be turned into classical states. We now make use of the results in the last three sections to quantify quantum correlations. We do so based on a recent protocol given in\cite{arXiv:1203.0011}. The protocol goes as the following: (i) Alice and Bob start with a quantum correlated state of two qubits. (ii) Using unitary transformations Alice encodes a message on her part of the state. (iii) Alice's state is sent to Bob, and he measures the two qubits. (iv)
Bob's challenge is to decode Alice's message. Bob cannot decode the whole message, but he will be able to decode more of the message if he can unitarily interact the two qubits and then make a measurement, compared with when he does not have the capabilities to interact the qubits. The difference in the information between the two cases turns out to be quantum discord. 

In this article we are interested in quantifying the cost of coherent interaction, i.e., the former case when Bob can interact two qubits. Bob can simply take the encoded state and apply a joint unitary operation such that the state becomes classical. We can measure the coherent-correlation content by taking measure function of the set of unitary operations that take entangled states to separable, or quantum to classical. Once a state is classical it can be measured locally and deduce the classical incoherent correlations. As we already showed that quantum coherent interaction, i.e., unitary operations, cannot take a density operator to a product state except in very limited cases. Therefore, the classical correlations contain the quantum-encoded information at this state. By doing this Bob has extracted the maximum amount of information about Alice's encoding.

%%%%%%%%%%%%%%%%%%%%%%%%%%%%%%%%%%%%%%%%
\subsection{Coherent correlations}
%%%%%%%%%%%%%%%%%%%%%%%%%%%%%%%%%%%%%%%%

Bob's ability to extract information out of a given state has little to do the state and more to do with its eigenbasis. Based on that intuition let us define a measure \emph{coherent correlation}

{\bf Definition.} Coherent correlation in a multipartite state are
\begin{gather}
\mathcal{D}(\rho \to \chi)=f(d_{AB})-\|U_{cl}\|.
\end{gather}
where $U_{cl}$ acting on the state $\rho$ gives a classical state and $\|\mathbb{I} \| = f(d_{AB})$. The function $f(d_{AB})$ makes sure that when $\rho$ is a classical state and $U_{cl}$ is the identity matrix then the measure of quantum-coherent interaction vanishes. If communication is allowed that Bob would simply make one of the two qubits classical and thereby spend less coherent resource. Then the measure quantum-coherent interaction would be
\begin{align}
\mathcal{D}(\rho \to \chi_{A|B}) &= f(d_{AB})- \max_{U^B_{cl}} \left\| U^B_{cl} \right\| \quad {\rm or}\\
	\mathcal{D}(\rho \to \chi_{B|A}) &= f(d_{AB})- \max_{U^A_{cl}} \left\| U^A_{cl} \right\|.
\end{align}
where $U^A_{cl}$ and $U^B_{cl}$ take Alice's and Bob's qubits to classical states respectively. This measure of quantum correlations here does not depend on the eigenvalues of the density operator. This means that state that is a mixture of Bell states has equivalent amount of quantum-coherent correlations to one pure Bell state. However the classical correlations of the two states will differ.

We have not specified the which norm would be appropriate on purpose as this would depend on the application. For sake of example let us consider $\mathcal{D}_2(\rho \to \chi)d_{AB}-\max_{U_{cl}} \left| \mbox{tr}(U_{cl}) \right|$. Given a quantum state $\rho=\sum_{ab} p_{ab} \ket{v_{ab}}\bra{v_{ab}}$, the coherent-quantum correlations are given by
\begin{gather}
\mathcal{D}(\rho \to \chi_{AB})=d_{AB}-\max_{\ket{ab}} \left| \sum_{ab} \braket{ab|v_{ab}} \right|
\end{gather}
for the symmetric case. When communication is allowed, the measure yields
\begin{align}
\mathcal{D}(\rho \to \chi_{B|A}) &=	d_{AB}- \max_{\ket{\phi_{a|b} \; b }} \left| \sum_{ab} \braket{\phi_{a|b} \;b |v_{ab}} \right|\\
\mathcal{D}(\rho \to \chi_{A|B}) &= d_{AB}-\max_{\ket{a \; \phi_{b|a}}} 	\left| \sum_{ab} \braket{a \; \phi_{b|a}|v_{ab}} \right|.
\end{align}
The quantity above measure the statistical distance between two basis, which can be interpreted as the distance along the shorted unitary path from a quantum state to a classical state. It is also a familiar quantity in many-body physics, when correlations are considered for adiabatic transformations are considered\cite{RevModPhys.80.517}.

%%%%%%%%%%%%%%%%%%%%%%%%%%%%%%%%%%%%%%%%
\subsection{Incoherent correlations}
%%%%%%%%%%%%%%%%%%%%%%%%%%%%%%%%%%%%%%%%

We can define classical correlations in a state $\rho$ by its eigenvalues $\{p_{ab}\}$. Again we can define $p_a = \sum_b p_{ab}$ and $p_b = \sum_a p_{ab}$. All of these quantities are independent of the eigenbasis of the state. 

{\bf Definition.} We define the \emph{incoherent correlation content} as
\begin{gather}
\mathcal{C}(\rho)=\sum_{ab} \left|p_{ab}-p_ap_b \right|.
\end{gather}
This is the trace distance from a joint probability distribution to the product of marginal distributions. This is a meaningful measure since after the unitary interaction, in the protocol discussed above, Bob will have a classical state whose correlations contain the information about Alice's encoding. By this definition classical correlations is related to degree of mixedness and mutual information.

In fact, we have defined coherent and incoherent correlations in a very different way here than normally done for quantum and classical correlations. Our coherent correlations are independent of the eigenvalues of the density operator and incoherent correlations are independent of the basis of the density operator. In separating them, we are unable to compare quantum correlations to classical correlations. Here they act as different degrees of freedom and cannot be meaningfully compared to each other. While quantum and classical correlations in the usual sense are important for resource states, coherent and incoherent correlations are needed in the decoding process and therefore just as important.

%%%%%%%%%%%%%%%%%%%%%%%%%%%%%%%%%%%%%%%%
\subsection{Entanglement}
%%%%%%%%%%%%%%%%%%%%%%%%%%%%%%%%%%%%%%%%

\begin{figure}[!t]
\begin{center}
\resizebox{8.09 cm}{4.5 cm}
{\includegraphics{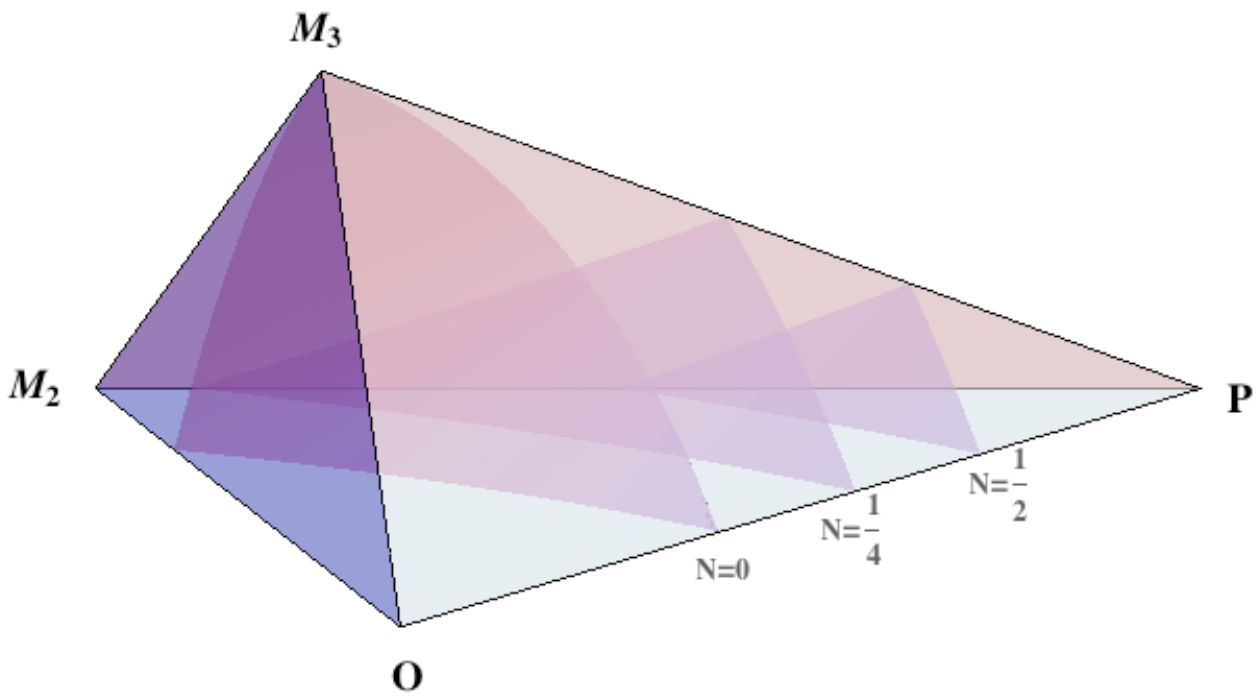}}
\caption{\label{entplane}(Color Online.) The three surface through the tetrahedron represent (from front to back) unitary orbits with maximum negativity of $0$, $\frac{1}{4}$, and $\frac{1}{2}$.}
\end{center}
\end{figure}

Lastly, let us discuss unitary orbits that contain states with entanglement briefly. Before we define measure for entanglement, we give a nice graphic illustration of entanglement in unitary orbits. To perform this analysis on equal footing we consider the maximum entanglement an orbit can posses\cite{verstraete} for two qubits. These states maximise negativity\cite{ppt}, concurrence\cite{hillwootters,wootterscon}, and relative entropy of entanglement\cite{PhysRevLett.78.2275,relentropy}. In Fig.~\ref{entplane} we have plotted equi-negativity surfaces within the tetrahedron. Note in the region near the fully mixed state there is no entanglement however almost all unitary orbits do not contain product states (see Fig.~\ref{productspace}).

Now we can define a measure of coherent-entanglement in this scenario very easily. Every entangled state $\rho$ can be mapped unitarily to a separable state $\sigma$ as $\sigma=U_s \rho U^\dag_s$. Then the measure of entanglement is give as
\begin{gather}
\mathcal{E}(\rho \to \sigma)=d_{AB}-\max_{U_s}\left| \mbox{tr}(U_s) \right|.
\end{gather}
Again, we are measuring the amount to coherent interaction needed to wash entanglement away, without disturbing the spectrum (and incidentally the classical correlations). This corresponds to the disentangling power of a unitary transformation\cite{PhysRevLett.103.030501}. 

Lastly, we can define coherent-dissonance\cite{arXiv:0911.5417} by considering the unitary transformations that map $\sigma$ to classical states:
\begin{gather}
\mathcal{Q}(\sigma \to \chi)=d_{AB}-\max_{U_{cl}}\left| \mbox{tr}(U_{cl}) \right|.
\end{gather}

Our motivations has been to quantify the coherent interaction needed to make a quantum state classical. This has a direct application for decoding information n terms of the protocol laid out in\cite{arXiv:1203.0011}. In this sense it may not be sensible to define entanglement and dissonance in the same spirit as quantum-coherent correlations. The measures defined here are not in the spirit of the general definition of quantum correlations either\cite{arXiv:1108.3649}. However they are meaningful in quantifying the amount of coherence needed to generate and decode correlations.

%%%%%%%%%%%%%%%%%%%%%%%%%%%%%%%%%%%%%%%%
%%%%%%%%%%%%%%%%%%%%%%%%%%%%%%%%%%%%%%%%
\section{Conclusion}
%%%%%%%%%%%%%%%%%%%%%%%%%%%%%%%%%%%%%%%%
%%%%%%%%%%%%%%%%%%%%%%%%%%%%%%%%%%%%%%%%
Correlations along a unitary orbit is not a well studied subject. This is partially because it is not particularly easy to tackle, see\cite{arXiv:1002.0314,arXiv:1110.2371,arXiv:1112.3372} for work extraction from correlations correlations along a unitary orbit as measured by mutual information\cite{PhysRevA.72.032317}. Using the notion of unitary orbits we have shown that all quantum states are unitarily connected to classical states, one-way or fully classical. We then we defined classical correlations in terms of the spectrum of the density operator and independent of the basis. This corresponds to classicality nicely as there is no notion of basis in classical probability theory. Quantum correlations are defined along a a unitary orbit by taking the shortest unitary path from a quantum state to classical state. Our measure of quantum correlation, in return does not depend on the eigenvalue of the density operator. These measure are defined in very different spirit to the typical approach. However, we interpreted both quantum and classical correlations in terms of a recent protocol. The quantum correlations are interpreted as the coherent interaction necessary to locally decode information contained in the state. In the same sprit as quantum-coherent correlations we defined entanglement and dissonance. Our approach here has not been rigorous, but we hope that it is found to be interesting and will generate further investigations.

\section*{Acknowledgments}
KM is supported by the John Templeton Foundation. KM and MG are financially supported by the National Research Foundation and the Ministry of Education of Singapore.

\section*{References}
\bibliographystyle{ws-ijmpb}
\bibliography{geo}
\end{document}